\documentclass[aps,pra,twocolumn,superscriptaddress,floatfix]{revtex4-2}
\usepackage{amsmath,amssymb,bm}
\usepackage{graphicx}
\usepackage{xcolor}
\usepackage{mathrsfs}
\usepackage{multirow}
\usepackage[colorlinks,linkcolor=magenta,citecolor=blue,urlcolor=blue]{hyperref}

\makeatletter
\newcommand{\Rmnum}[1]{\expandafter\@slowromancap\romannumeral #1@}
\makeatother

\begin{document}

\title{Impurity-induced loss bursts from anomalous scale-free localization in a non-Hermitian dissipative lattice}
%%%%%%%%authors%%%%%%%%%%%%%%%

\author{Hui Liu}
\affiliation{{Institute of Theoretical Physics and State Key Laboratory of Quantum Optics and Quantum Optics Devices, Shanxi University, Taiyuan, Shanxi 030006, China}}

\author{Zhihao Xu}
\email{xuzhihao@sxu.edu.cn}
\affiliation{{Institute of Theoretical Physics and State Key Laboratory of Quantum Optics and Quantum Optics Devices, Shanxi University, Taiyuan, Shanxi 030006, China}}

%%%%%%%%authors%%%%%%%%%%%%%%%

\date{\today}

\begin{abstract}
We identify anomalous scale-free localization and associated impurity-induced loss bursts in a non-Hermitian dissipative cross-stitch lattice. By a local basis rotation, the model is mapped onto an effective non-Hermitian Su-Schrieffer-Heeger lattice, where local impurities act as tunable effective boundaries. For the parameter choice considered here, the impurity-free periodic-boundary-condition point is reached at $\eta=1$. Tuning $\eta$ away from this point drives the system through generalized-boundary-condition regimes toward two effective open-boundary-condition-like limits at $\eta\to0$ and $\eta\to\infty$. For finite $\eta\notin\{0,1\}$, the loop-sector spectra remain separated from the real-energy axis, while the corresponding eigenstates exhibit scale-free localization pinned by the impurity. Unlike conventional impurity-induced scale-free localization, the Lyapunov exponent depends explicitly on the eigenenergy, making the localization strength eigenstate dependent. We further show that this anomalous eigenmode structure produces an impurity-induced loss burst: The long-time integrated dissipation probability develops a pronounced local enhancement relative to the surrounding background near an impurity-generated effective boundary, even when the initial wave packet is prepared far away. In the single-impurity case, the burst region consists of the impurity site and its adjacent effective-boundary site, and the effect occurs without imaginary-gap closing. For multiple impurities, local burst regions emerge around all impurities, while the dominant burst boundary is selected by the initial wave-packet position and the nonreciprocal drift direction. These results connect anomalous scale-free localization with controllable dissipation dynamics in non-Hermitian lattices.
\end{abstract}

\maketitle
		
\section{Introduction}

Spatial inhomogeneities play an essential role in non-Hermitian lattice systems. Unlike in Hermitian systems, where a local defect usually affects only a small number of bound states or scattering channels \cite{Plotnik2011,Hafezi2013}, a local perturbation in a non-Hermitian lattice can strongly reshape the global spectrum and eigenstates \cite{Guo2021,Li2020}. This sensitivity is closely related to the non-Hermitian skin effect, in which a macroscopic number of eigenstates accumulate near a boundary or an effective boundary generated by local perturbations \cite{Yao20181,Yao20182,Lee2019,Yang2020,Lee2020,Lee2021,Yoshida2020,Yi2020,Xiao2020,Lee2016,Li_2023,Yoshida2024,Zhang_2025,Kawabata2023,Gliozzi2024,Liu2024}. Local impurities can therefore provide a powerful way to control non-Hermitian spectra and to generate impurity-induced topological bound states \cite{Bosch2019,Liu2019,Liu2020}, disorder- or impurity-driven phase transitions \cite{Xu2022}, and non-Hermitian response phenomena \cite{Xiong2018,Kunst2018,Budich2020}.

A particularly intriguing impurity-induced phenomenon is scale-free localization (SFL) \cite{Guo2021,Li2021,Li2020,Yokomizo2021,Guo2023,Li2023,Molignini2023,Wang2023,Fu2023,Xie2024,Li2024,Liu2024b,Zhang2024,Spring2024,Zhang2025,Su2023,Yilmaz2024}. In contrast to ordinary skin states, whose localization length is typically independent of the system size, SFL states have localization lengths proportional to the system size. As a result, their eigenstate profiles collapse when plotted as functions of the normalized coordinate, while the finite-size spectra approach the periodic-boundary-condition (PBC) spectrum in the thermodynamic limit. In conventional impurity-induced SFL, the Lyapunov exponent is controlled by the impurity strength and the system size, and all eigenstates at a fixed impurity strength share the same localization tendency \cite{Guo2021,Li2021,Zhang2025}. The localization direction can be counterintuitive and may even be opposite to the bulk nonreciprocal directionality, but it remains common to the eigenstates governed by the same impurity parameter.

Dissipative non-Hermitian lattices also exhibit unusual dynamical responses. One representative example is the non-Hermitian edge burst \cite{Rudner2009,Xue2022,Xue2021,Qiao2024,Yuce2024,Zhu2024,Xiao2024,Wen2024,Hu2023,SLonghi2023,Ma2024}, where the long-time integrated loss probability develops an anomalously large peak near a boundary even when the initial wave packet is prepared far from that boundary. More recently, the non-Hermitian edge burst has been experimentally observed in photonic quantum-walk platforms \cite{Zhu2024,Xiao2024}, confirming that this anomalous boundary-loss enhancement is directly accessible in dynamical measurements. In the original edge-burst scenario, this effect was associated with the interplay between the non-Hermitian skin effect and imaginary-gap closing \cite{Xue2022,Ma2024}. Later studies showed that related burstlike responses can also appear in systems without conventional skin localization or under more general non-Hermitian dynamical conditions \cite{Yuce2023,Sen2025}. These developments raise a natural question: Can a local impurity in the bulk of a periodic dissipative lattice generate a loss burst analogous to a boundary edge burst, and, if so, what eigenstate structure supports it?

In this work, we address this question by considering a non-Hermitian dissipative cross-stitch lattice with tunable local impurities. By applying a local basis transformation, the system can be mapped onto an effective non-Hermitian Su-Schrieffer-Heeger (SSH) lattice \cite{Yao20181,Xue2022}, in which the impurities play the role of tunable effective boundaries. Varying the impurity strength drives the system between two effective open-boundary-condition (OBC)-like limits through generalized-boundary-condition (GBC) regimes and an impurity-free PBC point. This provides a controlled setting for studying how impurity-generated effective boundaries influence both eigenstate localization and dissipative dynamics.

We show that the loop-sector eigenstates in the finite-impurity regimes exhibit anomalous scale-free localization (ASFL). Their localization length still scales with the system size, as in conventional impurity-induced SFL, but the Lyapunov exponent depends explicitly on the eigenenergy through the impurity-modified transfer relation. Consequently, different eigenstates at the same impurity strength can have different localization strengths.

This anomalous eigenstate structure leads to a distinctive dissipative response. We find that an initially localized wave packet can produce pronounced long-time loss peaks near an impurity-generated effective boundary, even when the initial position is far from the impurity. This effect, which we call an impurity-induced loss burst, is characterized by the integrated dissipation probability on the lossy sublattice. For a single impurity, the burst region consists of the impurity site and the adjacent effective-boundary site. Notably, the burst occurs in regimes where the spectral loops remain separated from the real-energy axis, indicating that it is not triggered by imaginary-gap closing. For multiple impurities, each impurity generates a local burst region, while the dominant burst boundary is selected dynamically by the initial wave-packet position and the direction of nonreciprocal drift.

The remainder of this paper is organized as follows. In Sec.~\Rmnum{2} we introduce the dissipative cross-stitch lattice and its exact mapping to an effective non-Hermitian SSH model. In Sec.~\Rmnum{3} we analyze the single-impurity spectrum and derive the energy-dependent Lyapunov exponent characterizing ASFL states. In Sec.~\Rmnum{4} we study the single-impurity loss burst and clarify its distinction from the conventional edge burst. In Sec.~\Rmnum{5} we extend the analysis to multiple impurities and discuss the selection of the dominant burst boundary. Section Sec.~\Rmnum{6} summarizes the results.

\begin{figure*}[tbp]
  \includegraphics[width=0.95\textwidth]{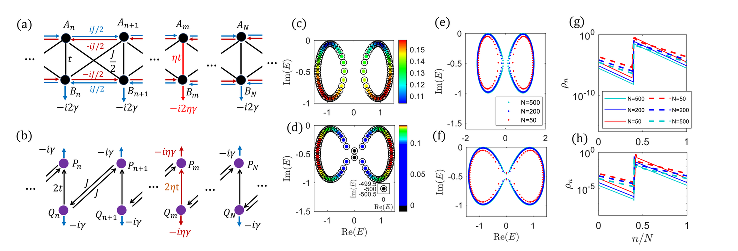}
  \caption{Schematic of the non-Hermitian impurity lattice, representative spectra, and energy-dependent eigenstate profiles. 
  (a) One-dimensional lossy cross-stitch lattice with local impurities of strength $\eta$. Same-sublattice intercell hoppings are direction-dependent imaginary hoppings with amplitudes $\pm iJ/2$, with opposite sign patterns on the $A$ and $B$ sublattices, while cross-sublattice intercell hoppings are real and equal to $J/2$.
  (b) Effective non-Hermitian SSH lattice obtained by the local rotation in Eq.~(\ref{eq3}), where the impurity modifies both the unidirectional intracell hopping and the onsite loss.
  Spectra of the single-impurity system are shown for(c) $\eta=10^{-3}$ and (d) $\eta=10^{3}$, with $N=50$. The color encodes the magnitude of the Lyapunov exponent, and the black circles denote the analytical roots of Eq.~(\ref{eq9}). The inset of (d) shows eigenvalues outside the plotting range of the main panel.
  Loop spectra are shown for a single impurity at $m=0.4N$ and different system sizes for (e) $\eta=10^{-3}$ and (f) $\eta=10^{3}$.
  (g) and (h) Unit-cell density distributions $\rho_n=|\psi_n^A|^2+|\psi_n^B|^2$ of selected normalized right eigenstates plotted versus $n/N$. 
  In (g) the dashed curves correspond to the eigenstates with the maximum imaginary part within the right spectral loop in (e), while the solid curves correspond to the eigenstates with the minimum imaginary part within the same loop.
  In (h) the solid curves correspond to the eigenstates with the maximum real part within the right spectral loop in (f), while the dashed curves correspond to selected eigenstates from the small-positive-real-part region of the same loop for different system sizes. The exact eigenvalues of all states used in (g) and (h) are listed in Appendix~\ref{app:state_selection}; these eigenvalues uniquely identify the plotted eigenstates. Here $J=1$ and $2t=2\gamma=1$.}
  \label{fig1}
\end{figure*}

\section{Model and exact mapping}

We consider a one-dimensional lossy cross-stitch lattice with impurities, as illustrated in Fig.~\ref{fig1}(a). The lattice contains $N$ unit cells, each consisting of two sublattice sites, labeled $A$ and $B$. Unless otherwise stated, periodic boundary conditions are imposed, i.e., $\psi_{N+1}^{A,B}=\psi_1^{A,B}$ and $\psi_{0}^{A,B}=\psi_N^{A,B}$. Impurities are introduced at the unit cells specified by $\mathcal{M}=\{m_1,m_2,\ldots,m_\kappa\}$, where $\kappa$ is the number of impurities and $n\in\mathcal{M}$ denotes an impurity cell.

The dynamical equations for the field amplitudes $\psi_n^A$ and $\psi_n^B$ are
\begin{align}
  i\frac{d\psi_n^A}{dt}= & t_n\psi_n^B+\frac{iJ}{2}(\psi_{n-1}^A-\psi_{n+1}^A)+\frac{J}{2}(\psi_{n-1}^B+\psi_{n+1}^B),\notag \\
  i\frac{d\psi_n^B}{dt}= & t_n\psi_n^A-\frac{iJ}{2}(\psi_{n-1}^B-\psi_{n+1}^B)+\frac{J}{2}(\psi_{n-1}^A+\psi_{n+1}^A)\notag  \\
                         & -i\gamma_n\psi_n^B .
  \label{eq1}
\end{align}
The intercell hopping pattern in Fig.~\ref{fig1}(a) follows Eq.~(\ref{eq1}): The same-sublattice hoppings are direction-dependent imaginary hoppings with amplitudes $\pm iJ/2$, with opposite sign patterns on the $A$ and $B$ sublattices, while the cross-sublattice hoppings are real and equal to $J/2$.

The intracell hopping amplitude $t_n$ and the loss rate $\gamma_n$ on the $B$ sublattice are defined as
\begin{align}
  t_n=
  \begin{cases}
    t, & n\notin\mathcal{M},\\
    \eta/2, & n\in\mathcal{M},
  \end{cases}
  \quad
  \gamma_n=
  \begin{cases}
    2\gamma, & n\notin\mathcal{M},\\
    \eta, & n\in\mathcal{M}.
  \end{cases}
  \label{eq2}
\end{align}
Here $t>0$ and $2\gamma>0$ are the bulk intracell hopping and loss rate, respectively, while $\eta\geq0$ specifies the impurity strength. Thus, an impurity cell has intracell hopping $\eta/2$ and loss rate $\eta$ on the $B$ sublattice.

To reveal the underlying structure, we perform the local basis rotation \cite{Yao20181,Xue2022}
\begin{align}
  U=\frac{1}{\sqrt{2}}\begin{pmatrix}  1&-i\\  -i&1  \end{pmatrix},
  \label{eq3}
\end{align}
which maps the cross-stitch lattice onto an effective non-Hermitian SSH lattice, as illustrated in Fig.~\ref{fig1}(b). For the case $t=\gamma$ used here, the mapped Hamiltonian becomes
\begin{align}
  H_{\mathrm{SSH}}= & \sum_{n\notin\mathcal{M}}\left(2t P_n^{\dagger}Q_n-i\gamma P_n^{\dagger}P_n-i\gamma Q_n^{\dagger}Q_n\right) \notag\\
  &+\sum_{n=1}^N J\left(P_{n+1}^{\dagger}Q_n+ Q_n^{\dagger}P_{n+1}\right) \notag\\
  &+\sum_{m_j\in\mathcal{M}}\left(\eta P_{m_j}^{\dagger}Q_{m_j}-i\frac{\eta}{2}P_{m_j}^{\dagger}P_{m_j}-i\frac{\eta}{2}Q_{m_j}^{\dagger}Q_{m_j}\right).
  \label{eq4}
\end{align}
Here $P_n$ and $Q_n$ denote the annihilation operators on the rotated sublattices. Under this rotation, the intercell hoppings in Eq.~(\ref{eq1}) become the reciprocal SSH intercell coupling $J$, whereas the nonreciprocity appears as asymmetric intracell hopping. In Fig.~\ref{fig1}(b), the impurities therefore correspond to local modulations of both the unidirectional intracell hopping and the onsite loss in the effective SSH lattice.

The parametrization in Eq.~(\ref{eq2}) is chosen to make this SSH representation simple. In general, a cell with intracell hopping $t_n$ and loss rate $\gamma_n$ is transformed into intracell hoppings $t_n+\gamma_n/2$ and $t_n-\gamma_n/2$, together with an onsite loss $-i\gamma_n/2$. For the symmetric bulk $t=\gamma$, this gives a unidirectional intracell hopping $2t$ and onsite loss $-i\gamma$. For an impurity cell, the choice of $t_m=\eta/2$ and $\gamma_m=\eta$ gives a unidirectional intracell hopping $\eta$ and onsite loss $-i\eta/2$. Hence the impurity preserves the same unidirectional SSH structure as the bulk while changing its local hopping and loss strengths.

The impurity cell becomes identical to the bulk cell only when $\eta/2=t$ and $\eta=2\gamma$ simultaneously. Therefore, within this single-parameter interpolation, the impurity-free point requires $t=\gamma$. In the following, we take $J=1$ and $2t=2\gamma=1$, for which the impurity-free PBC point is $\eta=1$. All energies are measured in units of $J$. In the opposite limits $\eta\to0$ and $\eta\to\infty$, transmission across the impurity region is effectively suppressed, so the impurity acts as an emergent boundary. Thus, tuning $\eta$ connects two effective OBC-like limits through finite-$\eta$ GBC regimes and the PBC point at $\eta=1$.

The qualitative phenomena discussed below do not rely on the fine tuning $t=\gamma$ or on the particular parametrization in Eq.~(\ref{eq2}). In Appendix~\ref{app:robust} we show that both ASFL and the impurity-induced loss burst persist for $t\neq\gamma$ and under an alternative impurity parametrization with a dimensionless scaling factor $s$.

\section{Single-impurity spectrum and ASFL}

In this section we analyze the single-impurity problem, where the impurity set reduces to $\mathcal{M}=\{m\}$. The exact mapping in Eq.~(\ref{eq4}) allows the single-impurity eigenvalue problem to be reduced to a scalar transfer problem in the effective SSH lattice. We denote by $p_n$ and $q_n$ the single-particle amplitudes on the $P$ and $Q$ sublattices, respectively. After eliminating $p_n$, the problem is reduced to a transfer equation for $q_n$. Away from the impurity, the transfer is governed by the impurity-free factor
\begin{align}
  q_{n}=\frac{E_1^2-J^2}{2tJ}q_{n-1},
  \label{eq6}
\end{align}
where $E_1=E+i\gamma$. The impurity replaces two consecutive impurity-free transfer steps by an impurity-modified two-step transfer from $q_{m-1}$ to $q_{m+1}$,
\begin{align}
   2t \eta J^2  q_{m+1}=(E_1E_2-J^2)^2 q_{m-1},
   \label{eq8}
\end{align}
with $E_2=E+i\eta/2$. The derivation is given in Appendix~\ref{app:single}. The key point is that the impurity-induced two-step transfer factor is explicitly energy dependent. Imposing the periodic closure condition around the ring gives the single-impurity self-consistency equation
\begin{align}
  (E_1E_2-J^2)^2(E_1^2-J^2)^{N-2} = \eta J(2tJ)^{N-1}.
  \label{eq9}
\end{align}
Equation~(\ref{eq9}) determines the single-impurity spectrum and shows how the impurity strength $\eta$ controls the global boundary closure of the effective SSH chain.

The spectral evolution can be understood directly from Eq.~(\ref{eq9}). In the limit $\eta=0$, transmission across the impurity cell is suppressed, and the impurity effectively cuts the ring into an OBC-like configuration. The spectrum then consists of two groups of highly degenerate eigenvalues: two $(N-2)$-fold degenerate levels at $E=\pm J-i\gamma$ and two additional doubly degenerate levels, $E=\frac{1}{2}\left(\pm\sqrt{4-\gamma^2}-i\gamma\right)$. When $\eta$ becomes finite, the degeneracies are lifted, and these degenerate levels broaden into a pair of symmetric spectral loops centered around $E=\pm J-i\gamma$. For $0<\eta<1$, the loops remain entirely in the lower half of the complex-energy plane and are separated from the real-energy axis. As $\eta$ increases, the loops expand continuously and eventually reach the PBC spectrum at $\eta=1$. At this point, the impurity modulation disappears, and Eq.~(\ref{eq9}) reduces to $E_1^2-J^2=e^{i2\pi l/N}$ for $l=0,1,\ldots,N-1$ so that the spectral loops become tangent to the real axis. 

For $\eta>1$, the spectrum moves away from the real-energy axis again, and an imaginary gap reopens between the spectral loops and the real axis. Meanwhile, four eigenvalues detach from the loop structure. The number of detached eigenvalues can be established by counting the asymptotic roots of Eq.~(\ref{eq9}). For a finite system, Eq.~(\ref{eq9}) is a polynomial equation of degree $2N$ in $E$. In the large-$\eta$ limit, the loop-sector roots are obtained from the solutions that remain close to $E_1^2-J^2=0$, namely $E=\pm J-i\gamma$. There are $2(N-2)$ such roots in total, with $N-2$ roots approaching each of the two OBC-like loop-sector levels. The remaining number of roots is therefore $2N-2(N-2)=4$, which defines the impurity sector. A large-$\eta$ expansion of Eq.~(\ref{eq9}) shows that two of these impurity-sector roots approach $E=-i\gamma$, while the other two approach $E=-i\eta/2$. These four roots are hybridized with the loop-sector roots at intermediate $\eta$, but they progressively separate from the spectral loops for large $\eta$ and become visible as detached eigenvalues. In the opposite limit $\eta\to\infty$, the system approaches another effective OBC-like regime. The two spectral loops shrink back to the two $(N-2)$-fold degenerate levels at $E=\pm J-i\gamma$, while the four detached eigenvalues form two doubly degenerate levels located at $E\to -i\gamma$ and $E\to -i\eta/2$. Therefore, although both $\eta=0$ and $\eta\to\infty$ correspond to effective OBC-like limits, the eigenvalues outside the loop sector are different in the two limits.

The analytical spectrum is confirmed by exact diagonalization. Figures~\ref{fig1}(c) and \ref{fig1}(d) show representative spectra obtained from exact diagonalization for $\eta=10^{-3}$ and $10^{3}$, respectively. The black circles denote the analytical roots obtained from the self-consistency equation in (\ref{eq9}), which are overlaid on the numerical spectra and coincide with the exact-diagonalization results. For $\eta=10^{3}$, the two impurity-sector eigenvalues approaching $E=-i\gamma$ remain within the plotting range of the main panel, whereas the other two eigenvalues approaching $E=-i\eta/2$ lie far below the loop spectrum and are shown in the inset of Fig.~\ref{fig1}(d). In the parameter regime considered here, the loop part of the spectrum remains separated from the real-energy axis for finite $\eta\notin\{0,1\}$, and the imaginary gap closes only when the system reaches the PBC point at $\eta=1$. For $\eta>1$, although four eigenvalues detach from the spectral loops, they also remain in the lower half of the complex-energy plane. Moreover, for any finite $\eta\notin\{0,1\}$, the spectral loops approach the PBC spectrum as the system size increases, as shown in Figs.~\ref{fig1}(e) and \ref{fig1}(f). This convergence of the finite-size spectrum toward the PBC spectrum is a characteristic spectral signature of scale-free localization \cite{Liu2024b,Zhang2025}.

The corresponding eigenstates provide further evidence for this scale-free behavior. Figures~\ref{fig1}(g) and \ref{fig1}(h) show unit-cell density distributions of representative eigenstates $\rho_n=|\psi_n^A|^2+|\psi_n^B|^2$ for different system sizes, plotted as functions of the normalized coordinate $n/N$. Since the transformation in Eq.~(\ref{eq3}) is local and unitary within each unit cell, this unit-cell density captures the same spatial scaling behavior as the amplitudes in the mapped SSH basis. The profiles collapse onto size-independent curves, consistent with a localization length that scales linearly with the system size. At the same time, the wave functions exhibit an abrupt change at the impurity position, reflecting the impurity-modified two-step recursion relation in Eq.~(\ref{eq8}). Thus, the impurity pins the spatial structure of the eigenstates, while the overall localization length remains scale-free.

A key observation is that the collapsed profiles are not identical for different eigenenergies. In Fig.~\ref{fig1}(g), for the $\eta=10^{-3}$ case, we compare two representative eigenstates selected from different regions of the same right spectral loop: The dashed curves correspond to the eigenstates with the maximum imaginary part, while the solid curves correspond to the eigenstates with the minimum imaginary part. 
In Fig.~\ref{fig1}(h), for the $\eta=10^{3}$ case, the solid curves correspond to the eigenstates with the maximum real part within the right spectral loop, while the dashed curves correspond to selected eigenstates from the small-positive-real-part region of the same loop for different system sizes. To make the state selection reproducible and unambiguous, the exact eigenvalues of all states plotted in Figs.~\ref{fig1}(g) and \ref{fig1}(h) are listed in Appendix~\ref{app:state_selection}. For both $\eta=10^{-3}$ and $\eta=10^{3}$, these representative states exhibit visibly different scaling profiles. Their relative spatial variation across the system, as well as their effective decay or growth trend away from the impurity, depends on the selected eigenenergy. Therefore, although these states display scale-free collapse, their localization properties cannot be characterized by the impurity strength alone.

To quantify this energy-dependent localization behavior, we introduce the Lyapunov exponent $\lambda$ through $|q_{n_{\mathrm{ref}}+L_d}|\approx e^{-\lambda L_d}|q_{n_{\mathrm{ref}}}|$, where $q_n$ denotes the eigenstate amplitude on the $Q$ sublattice of the $n$th unit cell and $L_d$ is the distance from a reference site $n_{\mathrm{ref}}$. A positive $\lambda$ indicates decay as the unit-cell index increases, a negative $\lambda$ indicates growth, and $\lambda=0$ corresponds to an extended profile. The corresponding localization length is given by $\xi=1/|\lambda|$. Combining the impurity-free recursion relation with the self-consistency equation, as detailed in Appendix~\ref{app:single}, we obtain
\begin{equation}
  \lambda(E,\eta,N)=\frac{1}{N}\left(\ln\left|\frac{2t}{\eta}\right|+
2\ln\left|\frac{E_1E_2-J^2}{E_1^2-J^2}\right|\right).
\label{eq11}
\end{equation}
This expression contains two contributions: the conventional impurity-controlled term and the energy-dependent term originating from the two-step transfer across the impurity region in Eq.~(\ref{eq8}). The latter accounts for the color variation along the spectral loops in Figs.~\ref{fig1}(c) and \ref{fig1}(d). Thus, even at fixed $\eta$ and $N$, different eigenstates generally possess different Lyapunov exponents.

Equation~(\ref{eq11}) provides a direct way to distinguish the present localized states from conventional impurity-induced scale-free localized states. In conventional SFL models, the impurity modifies the boundary closure condition through an energy-independent factor. In the notation of the present work, the corresponding Lyapunov exponent takes the form $\lambda_{\mathrm{conv}}=N^{-1}\ln\left|2t/\eta\right|$ \cite{Li2021,Zhang2025}. Thus, at fixed impurity strength, all eigenstates share the same localization tendency: The sign of $\lambda_{\mathrm{conv}}$ fixes the localization direction and its magnitude determines the localization length. Although changing $\eta$ across $2t$ can reverse the localization direction, the direction remains common to all eigenstates at the same $\eta$.

The present model is qualitatively different. As shown in Eq.~(\ref{eq8}), the impurity does not merely impose an energy-independent mismatch. Instead, the two-step transfer across the impurity contains the factor $(E_1E_2-J^2)^2$, which depends explicitly on the eigenenergy. This energy-dependent transfer factor is inherited by the Lyapunov exponent in Eq.~(\ref{eq11}). As a result, the localization strength becomes eigenstate dependent. The relative importance of the two terms in Eq.~(\ref{eq11}) depends on the impurity strength. For $\eta\ll1$, the conventional impurity-controlled term $\ln|2t/\eta|$ is large and dominates the Lyapunov exponent. By contrast, for $\eta\gg1$, the dominant contribution comes from the energy-dependent term $2\ln |(E_1E_2-J^2)/(E_1^2-J^2)|$, which reflects the impurity-modified two-step transfer across the impurity region. In both strongly asymmetric regimes, the dominant contribution fixes the sign of the Lyapunov exponent for the loop-sector eigenstates considered here, so that the corresponding wave functions exhibit a common skin tendency. When $\eta$ is close to the PBC value $\eta=1$, however, the impurity-controlled and energy-dependent contributions become comparable and can compete with each other. In this regime, the eigenenergy dependence can substantially modify the magnitude of the Lyapunov exponent and may even change its sign along the spectral loop.

Despite this energy dependence, the states identified here still retain the defining scale-free feature. For finite $\eta\notin\{0,1\}$, the Lyapunov exponent in Eq.~(\ref{eq11}) is proportional to $1/N$, so the localization length scales linearly with the system size, $\xi\sim N$. This accounts for the convergence of the finite-size spectra toward the PBC spectrum in the thermodynamic limit, while the eigenstates remain spatially nonuniform in finite systems. The distinction from conventional impurity-induced SFL therefore does not lie in the scale-free scaling itself, but in the energy-dependent coefficient of the $1/N$ Lyapunov exponent.

The limiting cases further clarify the connection with conventional SFL. At $\eta=1$, the impurity modulation vanishes and the eigenstates are extended. In the opposite limits $\eta=0$ and $\eta\to\infty$, the impurity suppresses transmission and produces OBC-like skin modes near the impurity-induced effective boundary. For finite $\eta\notin\{0,1\}$, the system interpolates between these limits and the PBC point through scale-free localized states. The distinctive feature here is that the coefficient of the $1/N$ Lyapunov exponent is energy dependent, making the localization strength eigenstate dependent. We refer to this unconventional form of scale-free localization as anomalous scale-free localization.

\section{Impurity-induced loss burst in the single-impurity system}

Here, by a loss burst we mean a nonlocal enhancement of the integrated dissipation probability: A pronounced loss peak appears at a site far from the initially occupied unit cell and is much larger than the surrounding background. In conventional non-Hermitian edge bursts, such an enhancement occurs at a physical boundary. As established in the preceding section, however, the loop-sector ASFL eigenstates are localized around the impurity with a localization length that scales with the system size. This scale-free localization, together with the role of the impurity as an effective boundary, provides the physical mechanism for the anomalous loss enhancement discussed below. We refer to this phenomenon as an impurity-induced loss burst.

\begin{figure}[tbp]
  \includegraphics[width=0.45\textwidth]{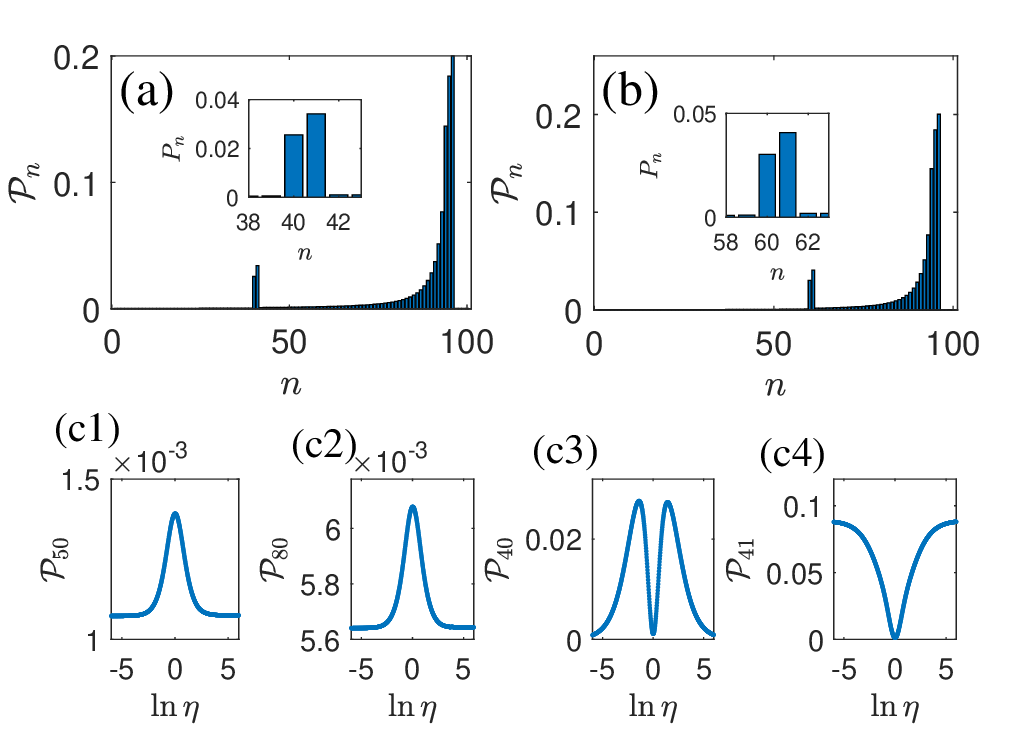}
  \caption{Single-impurity loss burst. The integrated dissipation probability $\mathcal{P}_n$ is plotted for a walker initially prepared at $n_0=95$, with the impurity located at (a) $m=40$ and (b) $m=60$, with the impurity strength fixed at $\eta=3$. The initial position is chosen to the right of the impurity so that the leftward nonreciprocal drift makes the wave packet encounter the impurity-generated effective boundary. Dependence of the dissipation probability on $\ln\eta$ for $m=40$ at four representative sites: (i) $n=50$, a nonburst site near the burst region; (ii) $n=80$, a site close to the initial position; (iii) $n=40=m$, the impurity site; and (iv) $n=41=m+1$, the adjacent effective-boundary site. The corresponding probabilities are denoted by $\mathcal{P}_{50}$, $\mathcal{P}_{80}$, $\mathcal{P}_{40}=\mathcal{P}_m$, and $\mathcal{P}_{41}=\mathcal{P}_{m+1}$, respectively. Here $N=100$, $J=1$, and $2t=2\gamma=1$.} \label{fig2}
\end{figure}

For the system described by Eq.~(\ref{eq1}) with a single impurity, the walker is initially prepared on sublattice $A$ of the $n_0$th unit cell, $\psi_n^A(0)=\delta_{n,n_0}$ and $\psi_n^B(0)=0$, which is normalized to unity. Since dissipation occurs only on sublattice $B$, the integrated loss provides a natural probe of the long-time absorption profile. In the parameter regimes considered below, the long-time survival probability vanishes within numerical accuracy. The corresponding dissipation probability at the $B$-sublattice site of the $n$th unit cell is defined as
\begin{align}
\mathcal{P}_n=2\int_0^{\infty}\gamma_n |\psi_n^B(t)|^2 dt,
\label{eq12}
\end{align}
which satisfies the normalization condition $\sum_n \mathcal{P}_n=1$ up to numerical precision. To quantify the burst response, we distinguish between the absolute burst-site probabilities and their contrast relative to the background. Here $\mathcal{P}_n$ denotes the integrated dissipation probability at unit cell $n$. For the single-impurity case, the absolute burst-site probabilities are given by $\mathcal{P}_m$ and $\mathcal{P}_{m+1}$, corresponding to the impurity site and its adjacent burst site, respectively. The burst-to-background contrasts are defined as $\mathcal{P}_m/\mathcal{P}_{\min}$ and $\mathcal{P}_{m+1}/\mathcal{P}_{\min}$, where $\mathcal{P}_{\min}$ denotes the minimum background dissipation probability between the impurity and the initial position of the wave packet. A finite-size scaling analysis based on these quantities is presented in Appendix~\ref{app:finite_size}. The analysis shows that the absolute probabilities at the burst sites decrease with increasing system size and hence do not remain a macroscopic fraction of the total dissipated probability in the thermodynamic limit. Nevertheless, the background minimum decreases faster with system size, so the burst-to-background contrast becomes increasingly pronounced.

The direction of the nonreciprocal drift determines the incident side of the impurity-generated boundary and hence the adjacent burst site. For the hopping convention used here, the wave packet drifts predominantly to the left. Therefore, when the initial wave packet is placed to the right of the impurity, it impinges on the impurity-generated boundary from the right, and the burst appears at the impurity site $m$ and the adjacent site $m+1$. If the Hamiltonian is kept unchanged but the initial wave packet is placed to the left of the impurity, the wave packet is initially driven away from the impurity. Under periodic boundary conditions, it can reach the impurity only after propagating around the ring, during which its probability weight is strongly attenuated by dissipation; the corresponding burst is therefore strongly suppressed. If the nonreciprocal hopping direction is reversed, the incident side of the impurity-generated boundary is also reversed. In particular, when the initial wave packet is placed to the left of the impurity and the drift is rightward, the wave packet directly impinges on the impurity from the left, and the corresponding burst region becomes $\{m,m-1\}$. A strictly mirror-related dissipation profile is obtained by additionally applying a spatial reflection, $n_0\rightarrow N+1-n_0$ and $m\rightarrow N+1-m$. Thus, the impurity-induced loss burst is not simply left-right symmetric for a fixed Hamiltonian; its position and strength are determined by the incident side selected jointly by the nonreciprocal drift direction and the initial wave-packet position.

Figure~\ref{fig2} shows that the burst persists for finite impurity strengths away from the PBC point, but in a modified form. For impurities located at $m=40$ and $60$ [Figs.~\ref{fig2}(a) and \ref{fig2}(b)], the dissipation probability develops pronounced local peaks at both the impurity site and the site immediately to its right, namely at $(n=40,41)$ and $(n=60,61)$, respectively. This identifies the burst region as $\mathcal{B}_m=\{m,\,m+1\}$. These results demonstrate that the anomalous enhancement is pinned not only to the impurity itself but also to its nearest-neighbor site on the incident side of the effective boundary, which is $m+1$ for the leftward drift considered here. We therefore refer to this phenomenon as an impurity-induced loss burst. In contrast to the conventional edge burst in the effective OBC-like limit, the present burst is not tied to a physical boundary of the lattice, but rather to the impurity-generated effective boundary in the bulk.

To further clarify the crossover from the finite-$\eta$ impurity-induced loss burst to the edge burst in the effective OBC-like limit, we examine the dependence of the dissipation probability on $\ln\eta$ at four representative sites for the case $m=40$: $n=50$, a non-burst site near the burst region; $n=80$, a site close to the initial position $n_0$; $n=40=m$, the impurity site; and $n=41=m+1$, the effective-boundary site. The corresponding probabilities are denoted by $\mathcal{P}_{50}$, $\mathcal{P}_{80}$, $\mathcal{P}_{40}=\mathcal{P}_{m}$, and $\mathcal{P}_{41}=\mathcal{P}_{m+1}$, respectively. The results are shown in Figs.~\ref{fig2}(i)--\ref{fig2}(iv). For sites away from the burst region, such as $n=50$ and $n=80$, the dissipation probability displays a pronounced Lorentzian-like dependence on $\ln\eta$, with a maximum at $\eta=1$, as shown by $\mathcal{P}_{50}$ and $\mathcal{P}_{80}$ in Figs.~\ref{fig2}(i) and \ref{fig2}(ii). As $|\ln\eta|$ increases away from zero, the dissipation probabilities at these non-burst sites decrease, indicating that more dissipation weight is redistributed toward the impurity region under the normalization condition $\sum_n \mathcal{P}_n=1$. Consequently, both $\mathcal{P}_m$ and $\mathcal{P}_{m+1}$ are enhanced. Despite this common enhancement, the subsequent evolution of $\mathcal{P}_m$ and $\mathcal{P}_{m+1}$ is qualitatively different. As $|\ln\eta|$ increases further, the impurity increasingly behaves as an effective open boundary and progressively suppresses transmission across the impurity region. During this crossover, $\mathcal{P}_m$ is initially enhanced by local scattering and trapping around the impurity. However, it eventually decreases when the dominant burst center shifts from the impurity site to its right neighboring site, $m+1$, leading to a bimodal dependence on $\ln\eta$. By contrast, the site $m+1$ continuously evolves into the conventional edge-burst site in the effective OBC-like limit, and therefore the corresponding dissipation probability $\mathcal{P}_{m+1}$ exhibits an inverse-Lorentzian-like behavior characteristic of a boundary burst. Here, an inverse-Lorentzian-like profile refers to a dependence with a minimum around $\ln\eta=0$ and enhanced values as $|\ln\eta|$ increases, opposite to the Lorentzian-like behavior observed at non-burst sites.

\begin{figure}[tbp]
  \includegraphics[width=0.45\textwidth]{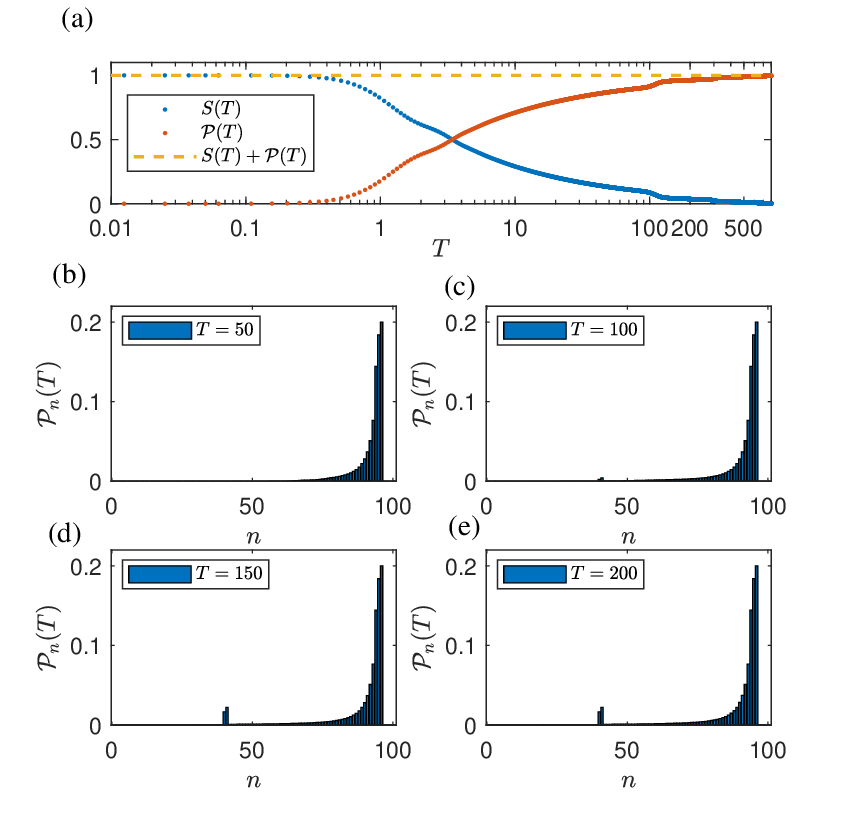}
  \caption{Finite-time signatures of the impurity-induced loss burst for $N=100$, $J=1$, $2t=2\gamma=1$, $\eta=3$, $m=40$, and $n_0=95$. 
(a) Survival probability $S(T)$, total dissipated probability $\mathcal{P}(T)$, and their sum $S(T)+\mathcal{P}(T)$ as functions of the observation time $T$.
Also shown are finite-time integrated dissipation profiles $\mathcal{P}_n(T)$ at (b) $T=50$, (c) $100$, (d) $150$, and (e) $200$.} 
  \label{fig7}
\end{figure}

The long-time results in Fig.~\ref{fig2} demonstrate that the impurity-induced loss burst appears in finite-$\eta$ regimes away from the PBC point, where the spectrum remains separated from the real-energy axis. Therefore, unlike the conventional non-Hermitian edge burst associated with imaginary-gap closing, the present burst is not triggered by such a spectral transition. Instead, it is a finite-impurity effect arising from the interplay between the impurity-generated effective boundary and the ASFL eigenstate structure. The effective boundary fixes the burst region around $m$ and $m+1$, while the ASFL states concentrate the long-time dissipation weight near this region, thereby producing the anomalously enhanced dissipation probability.

We also note that the loss burst can be detected within a finite observation time. Although Eq.~(\ref{eq12}) defines the long-time integrated dissipation probability, one may introduce the finite-time integrated profile
\begin{align}
\mathcal{P}_n(T)=2\int_0^{T}\gamma_n |\psi_n^B(t)|^2 dt .
\label{eq12b}
\end{align}
Such time-resolved loss distributions have already been experimentally accessed in photonic quantum-walk platforms, where spatially and temporally resolved photon-loss measurements allow the accumulated dissipation profile to be reconstructed \cite{Zhu2024,Xiao2024}. The corresponding total dissipated probability is given by $\mathcal{P}(T)=\sum_n\mathcal{P}_n(T)$, while the survival probability is
$S(T)=\sum_n\left(|\psi_n^A(T)|^2+|\psi_n^B(T)|^2\right)$. For the loss term $-i\gamma_n$ used in the Hamiltonian, the norm-balance relation reads
$dS(t)/dt=-2\sum_n\gamma_n|\psi_n^B(t)|^2$. Together with the normalized initial condition $S(0)=1$, this gives $S(T)+\mathcal{P}(T)=1$. As shown in Fig.~\ref{fig7}(a), $\mathcal{P}(T)$ increases with the observation time $T$, whereas $S(T)$ decreases, and their sum remains unity within numerical precision throughout the evolution. Figures~\ref{fig7}(b)--\ref{fig7}(e) display the finite-time integrated dissipation profiles $\mathcal{P}_n(T)$ at $T=50$, $100$, $150$, and $200$, respectively. At early times, the accumulated dissipation mainly follows the propagation path of the wave packet. Once the wave packet reaches the impurity-generated effective boundary, an additional localized peak emerges within the impurity-induced burst region. This impurity-associated peak is already visible at finite observation times and becomes increasingly pronounced as $T$ increases, while a finite survival probability still remains. Further increasing $T$ mainly drives the profile toward its long-time limit without appreciably shifting the burst position. Therefore, in photonic, acoustic, or circuit implementations with finite propagation length or finite measurement time, the loss burst can be identified through the finite-time emergence of enhanced absorption near the impurity-induced effective boundary.

\section{ASFL and loss-burst hierarchy in the multiple-impurity system}

Next, we extend the analysis to the case of multiple impurities. We denote the impurity set by $\mathcal{M}=\{m_1,m_2,\dots,m_\kappa\}$ and impose that no two impurities occupy nearest-neighbor unit cells, namely,
\begin{align}
d_{\mathrm{PBC}}(m_f,m_g)\ge 2,\qquad f\neq g, \notag
\end{align}
where $f,g\in\{1,2,\dots,\kappa\}$ and $d_{\mathrm{PBC}}(m_f,m_g)=\min\{|m_f-m_g|,\,N-|m_f-m_g|\}$ is the distance under periodic boundary conditions. This nearest-neighbor exclusion condition ensures that the local recursion relations associated with different impurities do not overlap, so that the impurity-modified two-step recursions can be applied independently.

\begin{figure*}[tbp]
  \includegraphics[width=0.9\textwidth]{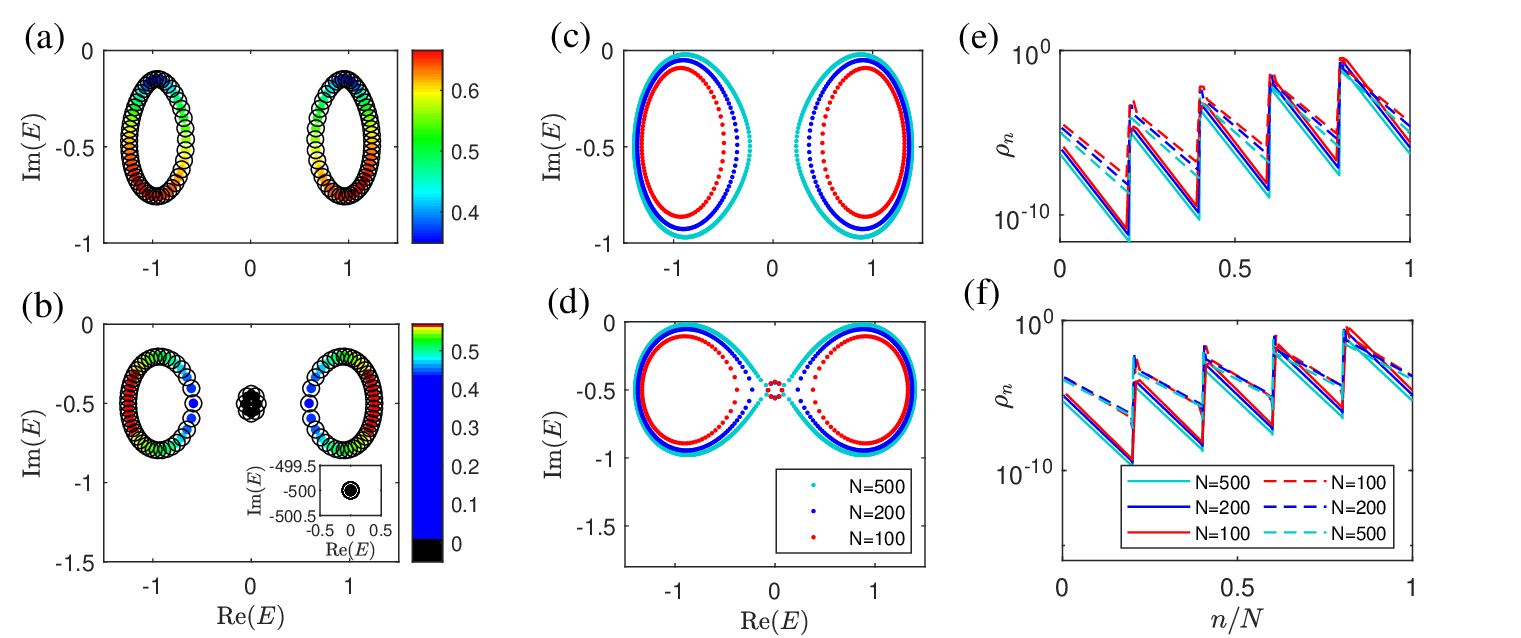}
  \caption{Spectra and energy-dependent scale-free eigenstate profiles of a system with four impurities located at $\mathcal{M}=\{0.2N,0.4N,0.6N,0.8N\}$. The system sizes are chosen such that all impurity positions are integers. 
   Spectra are shown for (a) $\eta=10^{-3}$ and (b) $\eta=10^{3}$, with $N=50$. The color encodes the magnitude of the multi-impurity Lyapunov exponent $\lambda_\kappa$. The analytical roots of Eq.~(\ref{eq13}) are shown as black circles. The inset of (b) shows the eigenvalues of the same spectrum that lie outside the plotting range of the main panel.
  Loop spectra are shown for (c) $\eta=10^{-3}$ and (d) $\eta=10^{3}$, for different system sizes. Only the loop part of the spectrum is displayed. 
  (e) and (f) Unit-cell density distributions $\rho_n=|\psi_n^A|^2+|\psi_n^B|^2$ of selected normalized right eigenstates for different system sizes, plotted as functions of the normalized coordinate $n/N$. 
  In (e) the dashed curves correspond to the eigenstates with the maximum imaginary part within the right spectral loop in (c), while the solid curves correspond to the eigenstates with the minimum imaginary part within the same loop. 
  In (f) the solid curves correspond to the eigenstates with the maximum real part within the right spectral loop in (d), while the dashed curves correspond to selected eigenstates from the small-positive-real-part region of the same loop for different system sizes. The exact eigenvalues of all states used in (e) and (f) are listed in Appendix~\ref{app:state_selection}; these eigenvalues uniquely identify the plotted eigenstates.} \label{fig3}
\end{figure*}

As in the single-impurity case, the lattice can be mapped, through the rotation matrix in Eq.~(\ref{eq3}), onto the effective non-Hermitian SSH model in Eq.~(\ref{eq4}). For impurity configurations satisfying the nearest-neighbor exclusion condition, the scalar transfer relations associated with different impurities do not overlap. Therefore, the global closure condition contains $N-2\kappa$ impurity-free transfer steps and $\kappa$ impurity-modified two-step transfers. This gives the multi-impurity self-consistency equation
\begin{align}
  (E_1E_2-J^2)^{2\kappa}(E_1^2-J^2)^{N-2\kappa}=\eta^\kappa J^N(2t)^{N-\kappa}.
  \label{eq13}
\end{align}
The derivation is given in Appendix~\ref{app:multi}. Equation~(\ref{eq13}) depends only on the number of impurities $\kappa$ and not explicitly on their positions, provided the nearest-neighbor exclusion condition above is satisfied. This position independence follows from the scalar recursion structure: The global closure condition only counts the total number of impurity-modified two-step transfers. The impurity positions nevertheless determine where the eigenstate profiles develop abrupt changes and therefore play an essential role in the dissipative burst dynamics discussed below.

The corresponding Lyapunov exponent is
\begin{equation}
  \lambda_\kappa(E,\eta,N)=\kappa\lambda(E,\eta,N),
  \label{eq14}
\end{equation}
where $\lambda(E,\eta,N)$ is the single-impurity Lyapunov exponent in Eq.~(\ref{eq11}). Equations~(\ref{eq13}) and (\ref{eq14}) show that, for impurity configurations satisfying the nearest-neighbor exclusion condition, tuning the impurity strength continuously drives the system from an effective OBC-like limit to a GBC regime, then to the PBC point, and finally back through another GBC regime to the opposite effective OBC-like limit. In the following we focus on a finite number of impurities, with $\kappa$ fixed as the system size is varied. In this case, $\lambda_\kappa\propto 1/N$, and the associated localization remains scale-free.

To illustrate the spectral and eigenstate evolution described above, we consider the case with $\kappa=4$ impurities placed at $\mathcal{M}=\{0.2N,\,0.4N,\,0.6N,\,0.8N\}$. Figure~\ref{fig3} shows the energy spectra obtained from exact diagonalization and the unit-cell density distributions $\rho_n$ of representative eigenstates for two impurity strengths, $\eta=10^{-3}$ and $10^{3}$. The black circles denote the analytical roots obtained from the self-consistency equation in (\ref{eq13}), which are overlaid on the exact-diagonalization spectra. Their agreement confirms the validity of the multi-impurity transfer analysis. For $\eta=10^{3}$, the impurity-sector eigenvalues outside the plotting range of the main panel are displayed in the inset of Fig.~\ref{fig3}(b). Figures \ref{fig3}(a), \ref{fig3}(c), and \ref{fig3}(e) correspond to $\eta=10^{-3}$, whereas Figs \ref{fig3}(b), \ref{fig3}(d), and \ref{fig3}(f) correspond to $\eta=10^{3}$. The spectra in Figs.~\ref{fig3}(a) and \ref{fig3}(b) are obtained for $N=50$, with the color encoding the magnitude of the multi-impurity Lyapunov exponent $\lambda_\kappa$. Figures~\ref{fig3}(c) and \ref{fig3}(d) display the finite-size evolution of the loop spectra, while Figs.~\ref{fig3}(e) and \ref{fig3}(f) show the density profiles of selected eigenstates for different system sizes as functions of the normalized coordinate $n/N$.

In the extreme limit $\eta\to 0$, the impurities effectively cut the ring into OBC-like segments. In this limit, the eigenvalues $E=\frac{1}{2}\left(\pm\sqrt{4-\gamma^2}-i\gamma\right)$ become $2\kappa$-fold degenerate for each branch. The remaining eigenvalues collapse into two $(N-2\kappa)$-fold degenerate levels at $E=\pm J-i\gamma$. The corresponding eigenstates are non-Hermitian skin states localized near the impurity-induced effective boundaries. For finite impurity strengths with $\eta\notin\{0,1\}$, the degeneracies are lifted and a pair of symmetric spectral loops centered around $E=\pm J-i\gamma$ emerges. The finite-size evolution in Figs.~\ref{fig3}(c) and \ref{fig3}(d) shows that these loop spectra approach the PBC spectral loops as the system size increases, consistent with the scale-free nature of the localization.

The associated loop-sector eigenstates are ASFL states localized around the impurity regions with system-size-dependent localization lengths. Compared with the single-impurity case, their spatial profiles exhibit multiple abrupt changes pinned at the impurity positions, while the overall envelopes retain the scale-free dependence on the normalized coordinate. Importantly, the collapsed profiles remain energy dependent in the multiple-impurity system. 

The unequal peak heights near different impurities in Figs.~\ref{fig3}(e) and \ref{fig3}(f) originate from the multiplicative transfer structure of the ASFL wave functions. Following the transfer-factor formulation derived in Appendix~\ref{app:single}, the right-eigenstate amplitudes are generated by successively applying the impurity-free transfer factor $A(E)$ and the impurity-modified two-step transfer factor $B(E,\eta)$ along the ring. Thus, although the spectrum in Eq.~(\ref{eq13}) depends only on the number of impurities under the nearest-neighbor exclusion condition, the eigenstate amplitudes still depend on the ordered sequence of impurity-free segments and impurity-modified transfer steps. For the configuration used in Fig.~\ref{fig3}, the impurity-separated segments are not all of the same length: The segment connecting the impurities at $0.8N$ and $0.2N$ across the periodic boundary has length $0.4N$, whereas the other adjacent separations are $0.2N$. For an impurity-free segment of length $L_j$ bounded by impurity-modified transfer steps, the accumulated factor is of the form $A(E)^{L_j-2}B(E,\eta)$. Since the Lyapunov exponent scales as $1/N$ while $L_j$ scales as $N$, these accumulated factors can generate finite relative weights between different impurity-separated segments in the normalized coordinate. This transfer-induced weighting gives rise to the observed spatial imbalance of the right-eigenstate density, including the enhanced peak near the impurity adjacent to the longer segment. This static imbalance can influence the local loss distribution by changing the eigenmode weight near different impurities, but it does not uniquely determine the hierarchy of the dynamical loss bursts. The latter also depends on how strongly the initial state excites individual eigenmodes through the biorthogonal overlaps and, more directly, on the nonreciprocal probability flux that reaches each impurity during the time evolution.

For $\eta=10^{-3}$, Fig.~\ref{fig3}(e) compares two eigenstates selected from different regions of the same right spectral loop: The dashed curves correspond to the eigenstates with the maximum imaginary part, while the solid curves correspond to the eigenstates with the minimum imaginary part. For $\eta=10^{3}$, Fig.~\ref{fig3}(f) compares eigenstates selected from different regions of the right spectral loop: for each system size, the solid curve corresponds to the eigenstate with the maximum real part, whereas the dashed curve corresponds to a selected eigenstate from the small-positive-real-part region of the same loop. To make the state selection reproducible and unambiguous, the exact eigenvalues of all states plotted in Figs.~\ref{fig3}(e) and \ref{fig3}(f) are listed in Appendix~\ref{app:state_selection}. These representative states exhibit distinct scale-free envelopes and different relative weights near the impurity-induced effective boundaries. This behavior is the multi-impurity counterpart of the energy-dependent ASFL discussed in the single-impurity case and follows from the energy-dependent Lyapunov exponent in Eq.~(\ref{eq14}).

At $\eta=1$, the system reaches the PBC point, where the impurity modulation is absent. The spectrum recovers the PBC structure, with two symmetric loops tangent to the real-energy axis, and the corresponding eigenstates are extended and nearly uniform over the entire lattice for all system sizes. In the opposite extreme limit $\eta\to\infty$, $2\kappa$ eigenvalues approach $E\to-i\gamma$, while another $2\kappa$ eigenvalues scale as $E\to -i\eta/2$. The remaining eigenvalues collapse into two $(N-2\kappa)$-fold degenerate levels at $E=\pm J-i\gamma$, corresponding to the loop-sector states in the effective OBC-like limit. Similar to the $\eta\to0$ limit, the corresponding eigenstates become skin states localized near the impurity-induced effective boundaries. Thus, both extreme limits $\eta\to 0$ and $\eta\to\infty$ correspond to effective OBC-like regimes, in which the eigenstates take the form of skin modes localized near the impurity-induced boundaries.
\begin{figure}[tbp]
  \includegraphics[width=0.5\textwidth]{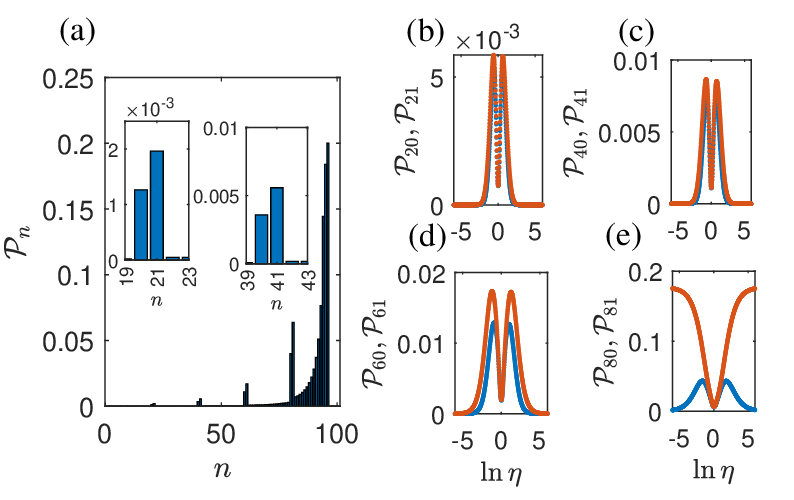}
  \caption{Multiple-impurity loss bursts. (a) Integrated dissipation probability $\mathcal{P}_n$ for a walker initially prepared at $n_0=95$ in a lattice with $N=100$ unit cells. Four impurities are located at $\mathcal{M}=\{20,40,60,80\}$. For the leftward nonreciprocal drift considered here, the impurity at $m=80$ is the first impurity encountered along the drift direction from the initial wave-packet position. Enlarged views of the weaker burst regions near the two leftmost impurities are shown in the insets. The impurity strength is fixed at $\ln{\eta}=1.4$, chosen in the intermediate GBC regime where local burst peaks at different impurities remain visible.  (b)--(e) Dependence of the paired burst probabilities $(\mathcal{P}_{m_j},\mathcal{P}_{m_j+1})$ on $\ln\eta$ for the four impurity positions $m_j\in\mathcal{M}$. Blue and orange dots correspond to $\mathcal{P}_{m_j}$ and $\mathcal{P}_{m_j+1}$, respectively.}\label{fig4}
\end{figure}

In a system with $\kappa$ impurities, the impurity-induced loss burst gives rise to $\kappa$ distinct burst regions. For the representative case with $\kappa=4$ considered here, four such regions can be clearly identified, with the burst region associated with the $j$th impurity given by $\mathcal{B}_{m_j}=\{m_j,\,m_j+1\}$ $(j=1,2,3,4)$. Figure~\ref{fig4}(a) shows the dissipation probability $\mathcal{P}_n$ for a walker initially prepared at $n_0=95$ for $N=100$ with $\ln{\eta}=1.4$. Although the dissipation probabilities at the left-side burst regions, such as $\mathcal{P}_{20}$, $\mathcal{P}_{21}$, $\mathcal{P}_{40}$, and $\mathcal{P}_{41}$, are much smaller than those at the rightmost burst region, $\mathcal{P}_{80}$ and $\mathcal{P}_{81}$, the insets show that they are still significantly enhanced relative to their neighboring nonburst sites. Thus, each impurity produces a local anomalous enhancement, while the overall magnitude of the burst response depends on the impurity position relative to the initial wave packet and the nonreciprocal drift direction.

Figures~\ref{fig4}(b)--\ref{fig4}(e) show the $\eta$ dependence of the burst-site probabilities. Since the initial wave packet is prepared at $n_0=95$ and the nonreciprocal drift is leftward, the impurity at $m_4=80$ is the first impurity encountered along the drift direction. As $|\ln\eta|$ increases, this impurity evolves into the dominant effective boundary. It both creates the local burst region $\mathcal{B}_{m_4}=\{80,81\}$ and suppresses transmission to the impurities on its left. Consequently, $\mathcal{P}_{81}$ retains the inverse-Lorentzian-like behavior of the single-impurity case, whereas $\mathcal{P}_{80}$ and the burst probabilities associated with $m=20,40,60$ exhibit bimodal profiles. These bimodal profiles reflect the competition between ASFL-induced redistribution toward impurity regions and the reduction of incoming probability flux caused by the first encountered impurity.

A clear spatial hierarchy appears in this crossover. Burst regions farther to the left require the wave packet to pass through more impurity-induced barriers and therefore lose their incoming flux earlier as $|\ln\eta|$ increases. For example, $\mathcal{P}_{20}$ starts to drop at $|\ln\eta|\approx0.6$, followed by $\mathcal{P}_{40}$ at $|\ln\eta|\approx0.68$, whereas $\mathcal{P}_{80}$ remains enhanced up to about $|\ln\eta|\approx1.7$. This hierarchy supports the picture that each impurity can host a local burst, while the dominant burst boundary is dynamically selected by the initial wave-packet position and the nonreciprocal drift direction.

\section{Conclusion}

In this work we have revealed an impurity-controlled route to anomalous scale-free localization and loss-burst dynamics in a non-Hermitian dissipative cross-stitch lattice. Through a local basis transformation, the model is exactly mapped onto an effective non-Hermitian SSH lattice, in which the impurity strength $\eta$ continuously tunes the system between two effective OBC-like limits reached at $\eta\to0$ and $\eta\to\infty$. Between these two limits, for the parameter choice used in this work, the system passes through generalized-boundary-condition regimes and reaches the PBC point at $\eta=1$. For finite $\eta\notin\{0,1\}$, the loop part of the spectrum remains separated from the real-energy axis, while the corresponding loop-sector eigenstates exhibit anomalous scale-free localization. Unlike conventional scale-free localized states, the Lyapunov exponent of the ASFL states depends explicitly on the eigenenergy, rendering the localization strength energy dependent.

The ASFL states lead to direct and observable dynamical consequences. In the single-impurity case, a walker released in the lattice develops pronounced long-time loss peaks at both the impurity site $m$ and the adjacent effective-boundary site $m+1$. The dissipation probability at $m+1$ follows an inverse-Lorentzian-like dependence on $\ln\eta$, whereas that at $m$ displays a bimodal profile. This distinguishes the impurity-induced loss burst from the conventional non-Hermitian edge burst associated with physical boundaries and, in the standard scenario, imaginary-gap closing. In the multiple-impurity configuration considered here, the nonreciprocal drift and the initial position dynamically select the dominant effective boundary. For the ordered configuration studied in this work, this boundary is the first impurity encountered along the drift direction, namely $m_\kappa$, and the dissipation probability at $m_\kappa+1$ retains the inverse-Lorentzian-like response. The remaining burst-site probabilities exhibit bimodal profiles due to the competition between ASFL-induced dissipation-weight redistribution and transmission suppression by impurity-generated effective boundaries.

Our results establish a direct connection between anomalous scale-free localization and impurity-induced loss-burst dynamics. They show that local impurity engineering can provide a flexible mechanism for controlling both eigenstate localization and long-time dissipation patterns in non-Hermitian dissipative lattices. The finite-size analysis further shows that, although the actual burst-site probabilities decrease with increasing system size, their enhancement relative to the background becomes more pronounced. In addition, the burst has a finite-time signature in the integrated dissipation profile, making it observable before the survival probability fully vanishes. These findings may be relevant to photonic waveguide arrays, acoustic resonator lattices, and electrical-circuit platforms, where engineered loss, nonreciprocal transport, and local impurities can be implemented and the time-resolved or integrated dissipation profile can be probed dynamically.

\begin{acknowledgments}
  Z. X. was supported by Quantum Science and Technology-National Science and Technology Major Project (Grant No. 2025ZD0300400), the NSFC (Grants No. 12375016 and No. 12461160324), and Beijing National Laboratory for Condensed Matter Physics (Grant No. 2023BNLCMPKF001).    
\end{acknowledgments}

\section*{Data Availability}
The data that support the findings of this article are available from the authors upon reasonable request.

\appendix

\section{Derivation of the single-impurity self-consistency equation and Lyapunov exponent}
\label{app:single}

In this appendix we derive the single-impurity self-consistency equation and the corresponding Lyapunov exponent. We consider the effective non-Hermitian SSH model in Eq.~(\ref{eq4}) with a single impurity at unit cell $m$. The eigenstate is written as
\begin{align}
  |\psi\rangle=\sum_n \left(p_n P_n^\dagger+q_n Q_n^\dagger\right)|0\rangle .
\end{align}
For impurity-free unit cells, the eigenvalue equations are
\begin{align}
   E_1 q_{n-1} &= Jp_n, \notag\\
   E_1 p_n &= Jq_{n-1}+2tq_n,
   \label{app_eq1}
\end{align}
where $E_1=E+i\gamma$. Eliminating $p_n$ gives
\begin{align}
  q_n=A(E)q_{n-1},\qquad
  A(E)=\frac{E_1^2-J^2}{2tJ}.
  \label{app_eq2}
\end{align}

The impurity modifies the local transfer relation. The eigenvalue equations involving the impurity cell $m$ and its right neighbor are
\begin{align}
  E_1 q_{m-1} &= J p_m, \notag\\
  E_2 p_m &= J q_{m-1}+\eta q_m, \notag\\
  E_2 q_m &= J p_{m+1}, \notag\\
  E_1 p_{m+1} &= J q_m+2t q_{m+1},
  \label{app_eq3}
\end{align}
where $E_2=E+i\eta/2$. Eliminating $p_m$ and $p_{m+1}$ yields
\begin{align}
  q_{m+1}=B(E,\eta)q_{m-1},\qquad
  B(E,\eta)=\frac{(E_1E_2-J^2)^2}{2t\eta J^2}.
  \label{app_eq4}
\end{align}
Thus, the impurity replaces two consecutive impurity-free transfer steps by one impurity-modified two-step transfer.

For a ring with a single impurity, the periodic closure condition is
\begin{align}
  A(E)^{N-2}B(E,\eta)=1.
  \label{app_eq5}
\end{align}
Substituting the explicit forms of $A(E)$ and $B(E,\eta)$ gives Eq.~(\ref{eq9}) in the main text.

We next derive the Lyapunov exponent. Away from the impurity,
\begin{align}
  |q_{n_{\mathrm{ref}}+L_d}|
  \simeq |A(E)|^{L_d}|q_{n_{\mathrm{ref}}}|.
\end{align}
Comparing this relation with
\begin{align}
  |q_{n_{\mathrm{ref}}+L_d}|
  \simeq e^{-\lambda L_d}|q_{n_{\mathrm{ref}}}|,
\end{align}
we obtain
\begin{align}
  \lambda=-\ln|A(E)|
  =
  \ln\left|
  \frac{2tJ}{E_1^2-J^2}
  \right|.\label{app_eq7}
\end{align}
Taking the logarithm of the modulus of Eq.~(\ref{eq9}) and rearranging terms gives
\begin{align}
  N\ln\left|
  \frac{2tJ}{E_1^2-J^2}
  \right|
  =
  \ln\left|\frac{2t}{\eta}\right|
  +2\ln\left|
  \frac{E_1E_2-J^2}{E_1^2-J^2}
  \right|.
\end{align}
Using Eq.~(\ref{app_eq7}), we finally obtain Eq.~(\ref{eq11})

\section{Derivation of the multi-impurity self-consistency equation}
\label{app:multi}

We now consider an impurity set
\begin{align}
  \mathcal{M}=\{m_1,m_2,\ldots,m_\kappa\},
\end{align}
with no two impurities occupying nearest-neighbor unit cells under periodic boundary conditions. This condition ensures that the impurity-modified two-step transfer regions do not overlap.

Away from the impurities, the transfer relation is the same as in Eq.~(\ref{eq6}),
\begin{align}
  q_n=\frac{E_1^2-J^2}{2tJ}q_{n-1}.
  \label{app_multi_eq1}
\end{align}
Each impurity contributes one two-step transfer,
\begin{align}
  q_{m_j+1}
  =
  \frac{(E_1E_2-J^2)^2}{2t\eta J^2}
  q_{m_j-1},
  \qquad m_j\in\mathcal{M}.
  \label{app_multi_eq2}
\end{align}
Thus, the global closure condition contains $N-2\kappa$ impurity-free steps and $\kappa$ impurity-modified two-step transfers:
\begin{align}
  \left(\frac{E_1^2-J^2}{2tJ}\right)^{N-2\kappa}
  \left[
  \frac{(E_1E_2-J^2)^2}{2t\eta J^2}
  \right]^\kappa
  =1,
  \label{app_multi_eq3}
\end{align}
or, equivalently, Eq.~(\ref{eq13}).

Taking the logarithm of the modulus of Eq.~(\ref{eq13}) gives
\begin{align}
  N\ln\left|
  \frac{2tJ}{E_1^2-J^2}
  \right|
  =
  \kappa\ln\left|\frac{2t}{\eta}\right|
  +2\kappa\ln\left|
  \frac{E_1E_2-J^2}{E_1^2-J^2}
  \right|.
\end{align}
Therefore, the multi-impurity Lyapunov exponent is
{\footnotesize
\begin{align}
  \lambda_\kappa(E,\eta,N)
  =
  \frac{\kappa}{N}
  \left[
  \ln\left|\frac{2t}{\eta}\right|
  +2\ln\left|
  \frac{E_1E_2-J^2}{E_1^2-J^2}
  \right|
  \right]
  =
  \kappa\lambda(E,\eta,N),
  \label{app_multi_eq5}
\end{align}}
which is Eq.~(\ref{eq14}). For fixed $\kappa$, $\lambda_\kappa\propto 1/N$, giving scale-free localization. 

\section{Eigenvalues of representative eigenstates}
\label{app:state_selection}

For reproducibility of the exact-diagonalization results, we list in Table~\ref{tab:selected_states_all} the eigenvalues of the representative normalized right eigenstates whose density profiles are plotted in Figs.~\ref{fig1}(g), \ref{fig1}(h), \ref{fig3}(e), and \ref{fig3}(f). These eigenvalues uniquely identify the plotted eigenstates for each system size and impurity configuration. All selected states belong to the right spectral loop of the corresponding finite-size spectrum. For the cases shown in Figs.~\ref{fig1}(g) and \ref{fig3}(e), which correspond to $\eta=10^{-3}$, the solid curves denote the eigenstates with the minimum imaginary part within the right spectral loop, whereas the dashed curves denote the eigenstates with the maximum imaginary part within the same loop. For the cases shown in Figs.~\ref{fig1}(h) and \ref{fig3}(f), which correspond to $\eta=10^{3}$, the solid curves denote the eigenstates with the maximum real part within the right spectral loop. The dashed curves denote representative eigenstates selected from the small-positive-real-part region of the right spectral loop for each system size.
The corresponding eigenvalues are listed in Table \ref{tab:selected_states_all} to make the state selection unambiguous.

\begin{table*}[tbp]
\caption{Eigenvalues of representative eigenstates used in the density-profile plots. The listed eigenvalues correspond to the solid and dashed curves in Figs.~\ref{fig1}(g), \ref{fig1}(h), \ref{fig3}(e), and \ref{fig3}(f).}
\label{tab:selected_states_all}
\begin{ruledtabular}
\begin{tabular}{ccccc}
Figure & $\eta$ & $N$ & Solid curve & Dashed curve \\
\hline
Fig.~\ref{fig1}(g) & $10^{-3}$ & 50  & $0.9048 - 0.9275i$ & $0.9046 - 0.0501i$ \\
Fig.~\ref{fig1}(g) & $10^{-3}$ & 200 & $0.8763 - 0.9810i$ & $ 0.8763 - 0.0135i$ \\
Fig.~\ref{fig1}(g) & $10^{-3}$ & 500 & $ 0.8702 - 0.9923i$ & $0.8702 - 0.0055i$ \\
Fig.~\ref{fig1}(h) & $10^{3}$ & 50  & $1.3705 - 0.5208i$ & $0.2932 - 0.6319i$ \\
Fig.~\ref{fig1}(h) & $10^{3}$ & 200 & $ 1.4032 - 0.5055i$ & $0.2300 - 0.6841i$ \\
Fig.~\ref{fig1}(h) & $10^{3}$ & 500 & $ 1.4098 - 0.5022i$ & $0.2329 - 0.7077i$ \\
Fig.~\ref{fig3}(e) & $10^{-3}$ & 100  & $0.9297 - 0.8638i$ & $ 0.9288 - 0.0906i$ \\
Fig.~\ref{fig3}(e) & $10^{-3}$ & 200 & $0.9048 - 0.9275i$ & $0.9046 - 0.0501i$ \\
Fig.~\ref{fig3}(e) & $10^{-3}$ & 500 & $0.8823 - 0.9699i$ & $0.8823 - 0.0212i$ \\
Fig.~\ref{fig3}(f) & $10^{3}$ & 100  & $1.3294 - 0.5000i$ & $0.4033 - 0.4071i$ \\
Fig.~\ref{fig3}(f) & $10^{3}$ & 200 & $1.3710 - 0.5000i$ & $0.3250 - 0.3312i$ \\
Fig.~\ref{fig3}(f) & $10^{3}$ & 500 & $1.3968 - 0.5000i$ & $0.2847 - 0.2863i$ \\
\end{tabular}
\end{ruledtabular}
\end{table*}

\section{Robustness of anomalous scale-free localization and impurity-induced loss bursts beyond the symmetric parameter choice}
\label{app:robust}

\begin{figure}[tbp]
  \includegraphics[width=0.5\textwidth]{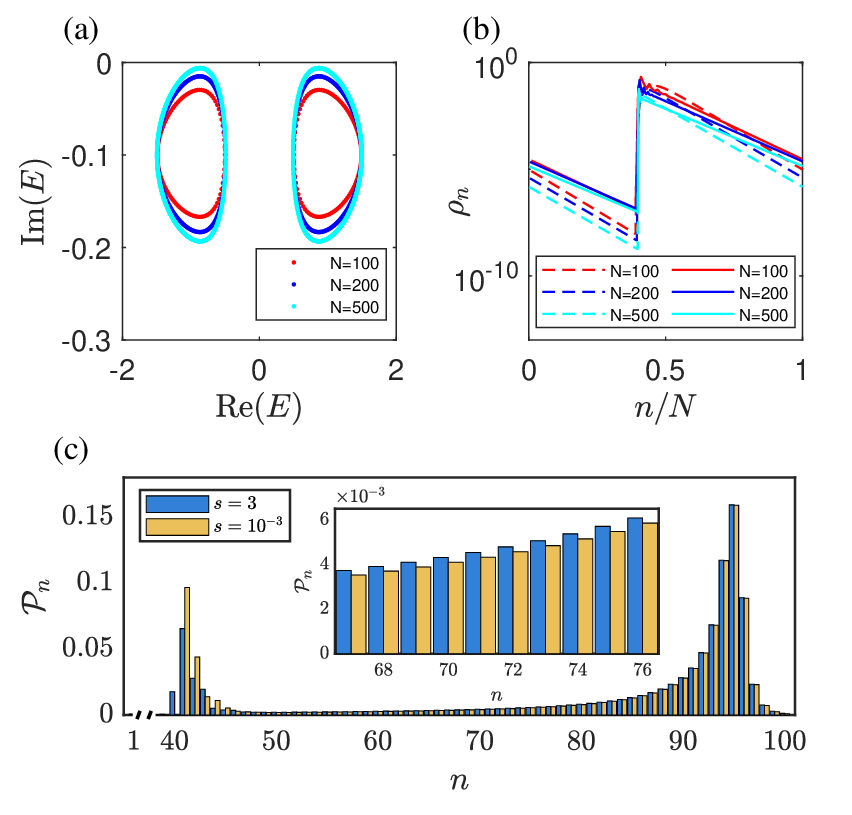}
  \caption{Robustness of ASFL and the impurity-induced loss burst beyond the symmetric condition $t=\gamma$ and under an alternative impurity parametrization. The parameters are $J=1$, $t=0.5$, $\gamma=0.1$, and the impurity position $m=0.4N$. The impurity strength is controlled by the dimensionless scaling factor $s$ defined in Eq.~(\ref{app_alt_param}).
(a) Complex-energy spectra for $s=10^{-3}$ and different system sizes $N=100$, $200$, and $500$.
(b) Probability-density profiles $\rho_n$ of representative right eigenstates for the same parameters as in (a). The solid curves correspond to the eigenstates with the smallest real part on the right spectral loop, while the dashed curves correspond to those with the largest real part on the right spectral loop. The horizontal coordinate is rescaled as $n/N$.
(c) Integrated dissipation probability $\mathcal{P}_n$ for $s=3$ and $10^{-3}$ in a lattice with $N=100$, $m=40$, and initial wave-packet position $n_0=95$. The inset enlarges a representative bulk region, showing that the bulk dissipation probability is suppressed as $|\ln s|$ increases.}
\label{figc1}
\end{figure}

In this appendix we examine whether the ASFL eigenstates and the impurity-induced loss burst rely on the symmetric parameter choice $t=\gamma$ or on the particular impurity parametrization used in the main text. To this end, we consider an alternative parametrization in which the impurity cell is introduced by multiplying both the intracell hopping amplitude and the local loss rate by the same dimensionless factor $s$:
\begin{align}
t_m=st,\qquad \gamma_m=2s\gamma .
\label{app_alt_param}
\end{align}
Here $s=1$ corresponds to the impurity-free lattice for arbitrary values of $t$ and $\gamma$.

After the local basis rotation in Eq.~(\ref{eq3}), the dissipative cross-stitch lattice is mapped to the following non-Hermitian SSH Hamiltonian:
{\footnotesize
\begin{align}
H_{\mathrm{SSH}}=&\, \sum_{n\notin\mathcal{M}}\left[(t+\gamma)P_n^\dagger Q_n
+(t-\gamma)Q_n^\dagger P_n
-i\gamma(P_n^\dagger P_n+Q_n^\dagger Q_n)\right]\notag\\
&+\sum_{n=1}^N J\left(P_{n+1}^\dagger Q_n+Q_n^\dagger P_{n+1}\right)\notag\\
&+\sum_{n\in\mathcal{M}}s\left[(t+\gamma)P_n^\dagger Q_n
+(t-\gamma)Q_n^\dagger P_n\right]\notag\\
&-\sum_{n\in\mathcal{M}}is\gamma
\left(P_n^\dagger P_n+Q_n^\dagger Q_n\right).
\label{app_c1}
\end{align}}
For $t\neq\gamma$, the intracell hopping in the SSH representation remains nonreciprocal but becomes bidirectional, with hopping amplitudes $t+\gamma$ and $t-\gamma$. In the original dissipative cross-stitch lattice, the impurity cell has intracell hopping amplitude $st$ and loss rate $2s\gamma$. In the numerical examples below, we choose $J=1$, $t=0.5$, and $\gamma=0.1$, which are far from the symmetric condition $t=\gamma$.

Figure~\ref{figc1}(a) shows the complex-energy spectra for $s=10^{-3}$ and several system sizes. The spectra exhibit a clear finite-size evolution toward the corresponding impurity-free PBC loop structure, indicating that the scale-free spectral behavior is not restricted to the symmetric point. The representative right-eigenstate profiles in Fig.~\ref{figc1}(b), plotted against the normalized coordinate $n/N$, further collapse onto smooth envelopes for different system sizes. The two energy-selected families separately collapse onto size-independent envelopes, while the solid and dashed families exhibit visibly different envelopes. This simultaneously demonstrates the scale-free scaling of the localization length and its eigenstate dependence, showing that the defining phenomenology of ASFL persists away from the symmetric condition $t=\gamma$.

We further examine the integrated dissipation profiles in Fig.~\ref{figc1}(c). For the moderate impurity strength $s=3$, the enhanced dissipation remains concentrated near the impurity-generated boundary, mainly around the impurity site $m$ and the incident-side neighboring site $m+1$, although the detailed spatial profile is quantitatively modified compared with the symmetric case discussed in the main text. For a stronger deviation from the impurity-free point, represented by $s=10^{-3}$, the impurity-associated dissipation is redistributed over the impurity site, the incident-side neighbor, and several nearby sites to the right. At the same time, the inset shows that the dissipation probability in the bulk region is suppressed as $|\ln s|$ increases, consistent with the suppression of non-burst-site dissipation with increasing impurity-strength deviation shown in Fig.~\ref{fig2} of the main text.

These results demonstrate that both key phenomena discussed in the main text are robust: the ASFL eigenstates survive beyond the symmetric condition $t=\gamma$, and the impurity-induced loss burst remains observable under an alternative impurity parametrization. The precise spatial shape and width of the burst may change quantitatively, but the impurity-induced redistribution of dissipation and the accompanying suppression of bulk dissipation remain qualitatively unchanged.

\section{Finite-size scaling of the loss-burst strength}
\label{app:finite_size}

\begin{figure}[tbp]
  \includegraphics[width=0.5\textwidth]{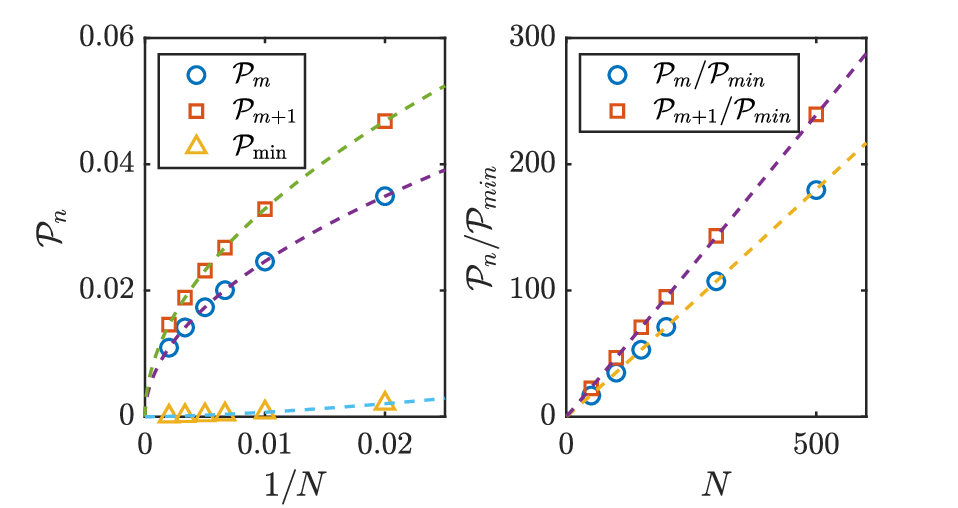}
  \caption{Finite-size scaling of the impurity-induced loss burst for $J=1$, $t=\gamma=0.5$, and $\eta=3$. The impurity and initial wave-packet positions are fixed at $m=0.2N$ and $n_0=0.9N$, respectively, while the system size is varied from $N=50$ to $500$. 
(a) Absolute burst-site dissipation probabilities $\mathcal{P}_m$ and $\mathcal{P}_{m+1}$, together with the minimum background probability $\mathcal{P}_{\min}$, plotted as functions of $1/N$. The dashed curves are power-law fits in $N$: $\mathcal{P}_m=0.2519 N^{-0.505}$, $\mathcal{P}_{m+1}=0.3409 N^{-0.5075}$, and $\mathcal{P}_{\min}=0.7887 N^{-1.519}$. 
(b) Burst-to-background contrasts $\mathcal{P}_m/\mathcal{P}_{\min}$ and $\mathcal{P}_{m+1}/\mathcal{P}_{\min}$ as functions of $N$. The dashed curves denote the corresponding fits: $\mathcal{P}_m/\mathcal{P}_{\min}=0.3283 N^{1.015}$ and $\mathcal{P}_{m+1}/\mathcal{P}_{\min}=0.4412 N^{1.013}$.}
  \label{figd1}
\end{figure}

In this appendix we quantify the finite-size behavior of the impurity-induced loss burst in terms of both the absolute burst-site probabilities and their relative enhancement over the background. The system contains $N$ unit cells, with a single impurity located at the unit cell $m$ and the initial wave packet placed at $n_0$. The parameters $J$, $t$, $\gamma$, and $\eta$ are defined in the main text.

The quantity $\mathcal{P}_n$ denotes the integrated dissipation probability at unit cell $n$, obtained by integrating the time-resolved loss probability over the full time evolution. With the normalization used in this work, the total integrated dissipation probability satisfies $\sum_n \mathcal{P}_n=1$, so $\mathcal{P}_n$ measures the fraction of the total dissipated probability absorbed at unit cell $n$. In the single-impurity geometry considered here, the loss burst appears at the impurity site $m$ and its adjacent site $m+1$ on the incident side of the nonreciprocal drift. We therefore use $\mathcal{P}_m$ and $\mathcal{P}_{m+1}$ as the absolute burst-site dissipation probabilities.

To characterize the burst contrast relative to the surrounding background, for the geometry considered here with $m<n_0$, we define
\begin{equation}
\mathcal{P}_{\min}=\min_{m\leq n\leq n_0}\mathcal{P}_n,
\end{equation}
where the minimization is taken over the propagation region between the impurity position $m$ and the initial wave-packet position $n_0$. The dimensionless ratios $\mathcal{P}_m/\mathcal{P}_{\min}$ and $\mathcal{P}_{m+1}/\mathcal{P}_{\min}$ are then used as burst-to-background contrasts. They measure the contrast between the burst-site dissipation and the weakest background dissipation in the propagation region.

Figure~\ref{figd1} shows the finite-size scaling for $J=1$, $t=\gamma=0.5$, and $\eta=3$, with $m=0.2N$ and $n_0=0.9N$. In Fig.~\ref{figd1}(a), $\mathcal{P}_m$, $\mathcal{P}_{m+1}$, and $\mathcal{P}_{\min}$ are fitted by the power-law form $y\simeq a_1 N^{-\alpha}$, where $a_1$ is a fitting prefactor and $\alpha$ is the finite-size scaling exponent. The fitted results show that the two burst-site probabilities decay as $\mathcal{P}_m=0.2519 N^{-0.505}$ and $\mathcal{P}_{m+1}=0.3409 N^{-0.5075}$, whereas the background minimum decays faster as $\mathcal{P}_{\min}=0.7887 N^{-1.519}$.

Therefore, the absolute burst-site probabilities $\mathcal{P}_m$ and $\mathcal{P}_{m+1}$ vanish algebraically in the large-system limit and do not remain a macroscopic fraction of the total dissipated probability. This behavior is consistent with the normalization condition $\sum_n \mathcal{P}_n=1$ and the scale-free nature of the spatial distribution. Nevertheless, because the background minimum decreases faster with $N$, the burst-to-background contrasts increase approximately linearly with system size, as shown in Fig.~\ref{figd1}(b). Thus, the impurity-induced loss burst becomes increasingly pronounced relative to the bulk background, even though the absolute probability at each burst site decreases in the thermodynamic limit.

\end{document}